\documentclass[pra,aps,twocolumn,superscriptaddress]{revtex4}
\usepackage{hyperref}
\usepackage{bm}
\usepackage{epsfig}
\usepackage{amssymb}
\usepackage{graphicx}

\usepackage{amsmath}
\usepackage{epsfig}

\begin{document}
\title{Continuous-variable teleportation of a negative Wigner function}
\author{Ladislav Mi\v{s}ta, Jr.}
\affiliation{Department of Optics, Palack\' y University, 17.
listopadu 12,  771~46 Olomouc, Czech Republic}
\author{Radim Filip}
\affiliation{Department of Optics, Palack\' y University, 17.
listopadu 12,  771~46 Olomouc, Czech Republic}
\author{Akira Furusawa}
\affiliation{Department of Applied Physics, University of Tokyo, 7-3-1 Hongo, Bunkyo-ku, Tokyo 113-8656, Japan}
\date{\today}

\begin{abstract}
Teleportation is a basic primitive for quantum communication and
quantum computing. We address the problem of continuous-variable
(unconditional and conditional) teleportation of a pure
single-photon state and a mixed attenuated single-photon state
generally in a nonunity gain regime. Our figure of merit is the
maximum of negativity of the Wigner function that witnesses highly
non-classical feature of the teleported state. We find that
negativity of the Wigner function of the single-photon state can
be {\em unconditionally} teleported for arbitrarily weak squeezed
state used to create the entangled state shared in the
teleportation. In contrast, for the attenuated single-photon state
there is a strict threshold squeezing one has to surpass in order
to successfully teleport the negativity of its Wigner function.
The {\em conditional} teleportation allows to approach perfect
transmission of the single photon for an arbitrarily low squeezing
at a cost of a success rate. On the other hand, for the attenuated
single photon conditional teleportation cannot overcome the
squeezing threshold of the unconditional teleportation and it
approaches negativity of the input state only if the squeezing
simultaneously increases. However, as soon as the threshold
squeezing is surpassed the conditional teleportation still
pronouncedly outperforms the unconditional one. The main
consequences for quantum communication and quantum
computing with continuous variables are discussed.
\end{abstract}

\maketitle
\section{Introduction}

Quantum teleportation is a fundamental primitive in quantum
information \cite{Bennett_93, Vaidman_94,Braunstein_98}.
Principally, it allows to remotely decompose quantum state to a
noise and necessary classical information required to uncover that
state from the noise. In the quantum key distribution, the
teleportation between distant parties combined with quantum
repeaters can transmit fragile quantum resources over a large
distance \cite{Briegel_98,Duan_01,Eisert_04}. In quantum
computation, the teleportation allows for fault-tolerant
deterministic implementation of a difficult quantum gate on an unknown quantum
state \cite{Gottesman_99, Bartlett_03}.

For quantum states in an infinite-dimensional Hilbert space,
quantum squeezing is an irreducible resource for the universal
quantum teleportation \cite{Vaidman_94,Braunstein_98}. Quantum
squeezed states are states with variance of a quadrature below the
vacuum noise. Although the squeezed states are non-classical, as
they have no regular and positive Glauber-Sudarshan
quasi-probability distribution, their non-classicality still can
be simulated by the semiclassical methods. It results from the
fact that squeezing is simply observable in a Gaussian
approximation, since the squeezed states are there represented by
the positive and regular Wigner functions which then play the role
of probability distributions. Such the Wigner function of a
squeezed state can be simply obtained by deforming the stochastic
phase-space evolution of an irreducible vacuum state
\cite{Kinsler_91,Filip_01,Olsen_09}. It does not prevent, for
example, quantum key distribution for a limited distance, but it
does not allow universal quantum computing \cite{Bartlett_02}. On
the other hand, non-classicality substantially reflecting a
discrete particle structure of quantum states cannot be
efficiently simulated by these stochastic methods \cite{Olsen_09}.
The corresponding Wigner function of the particle-like state can
exhibit negative values, breaking its interpretation as any kind
of classical probability density. The negative values are
considered as a clear experimental witness of quantum features
beyond the semiclassical description \cite{neg}.

Both the long-distance quantum key distribution and quantum
computing cannot be performed solely based on the squeezing
resource. The highly non-classical repeaters in the sequential
teleportation protocol \cite{Briegel_98,Duan_01,Eisert_04} or many
highly non-classical cubic phase gates \cite{Gottesman_01} in a
complex quantum computer change propagating states to non-Gaussian
states with a negative Wigner function. In both the cases, the
teleportation of the negative Wigner function with just the squeezing
as a resource is a basic element of the communication and
computation tasks. In the efficient quantum key distribution
with quantum repeaters, the Gaussian teleportation should,
at least {\em probabilistically}, allow to propagate the
negativity of the Wigner function produced by the repeater
operation through the network towards the next quantum repeater.
In quantum computation, it should even {\em deterministically}
implement basic highly nonlinear cubic phase gate
\cite{Gottesman_01}, if such the off-line gate is, at least
probabilistically, feasible.

The quality of teleportation is mainly limited by the finite
squeezing resource. How much squeezing resource is actually
required to at least partially keep the negativity of the Wigner
function through the teleportation step? The answer to this
question tells how much of the squeezing is necessary to decompose
higher non-classical states. Since the teleportation is the basic
primitive for the quantum communication and computation, it
specifies also an amount of the squeezing needed to
deterministically (or probabilistically) operate highly
non-classical states. In this paper, we give a clear and
illustrative answer to this basic question. As a first testing
state, the single-photon Fock state having maximal possible
negativity of the Wigner function is considered at the input of
teleportation. The negativity is then lowered by a loss
implemented on the single-photon state. Our attempt is to directly
show the effect of the Gaussian teleportation on the different
values of negativity of the Wigner function, thus judging its
possible application in the long-distance quantum communication
and quantum cryptography.

Previously, the mechanism of teleportation of a single-photon
state (or superposition of the coherent states) was analyzed in
Refs.~\cite{Braunstein_98,Lee_00,Hofmann_01,Kurzeja_02}, but
always using the fidelity \cite{Ide_02_ng,Chizhov_02}, entanglement fidelity
\cite{Johnson_02} or photon-number distribution \cite{Ide_02}, as
a figure of merit. For the unity gain teleportation, the fidelity
benchmark $2/3$ was found to be a necessary condition to achieve
teleportation of a negative Wigner function \cite{Ban_Caves_04}.
It corresponds to squeezing -3 dB required in the entanglement
preparation. However, fidelity does not directly tell if the
teleported state has still some negativity of Wigner function and
how large it is.

In this paper we investigate the capability of the standard
teleportation protocol \cite{Braunstein_98} to successfully
teleport negativity of the Wigner function in the origin of the
single-photon Fock state and the convex mixture of the state with
the vacuum state. Our goal is to find teleportation protocols
minimizing the value of the output Wigner in the origin. We find
that for the single-photon Fock state an arbitrarily small nonzero
squeezing suffices to successfully teleport negative value of the
Wigner function in the origin if the gain of teleportation is
chosen suitably. In contrast, in order to teleport negative value
of the Wigner function in the origin of the attenuated
single-photon Fock state one has to surpass a
strict threshold level of squeezing. For both the cases of input
states one can attain substantially larger negative value of the
output Wigner function in the origin by using conditional
teleportation with reasonably high success probability.
The post-selection cannot improve the squeezing threshold,
however, if it is surpassed, a higher negativity of the Wigner
function can be achieved. A sufficient tolerance of the
conditional teleportation to impurity of the squeezed states
used to produce the shared entangled state demonstrates the
feasibility of the conditional teleportation.

The paper is organized as follows. Sec.~\ref{sec_1} deals with
unity gain, optimal nonunity gain and conditional teleportation
of a single-photon and squeezed single-photon Fock state. In
Sec.~\ref{sec_2} we study unity gain, optimal nonunity gain and
conditional teleportation of a convex mixture of a single-photon
Fock state and the vacuum state. Sec.~\ref{sec_3} contains
conclusion.

\section{Teleportation of a single-photon Fock state}\label{sec_1}

At the outset we focus on understanding of the basic effects of CV
teleportation on the negativity of Wigner function. For this
purpose we start with the simple case of teleportation of a
single-photon Fock state $|1\rangle$. The state is described by
the following Wigner function \cite{Leonhardt_97}:
\begin{eqnarray}\label{Win}
W_{\rm in}(r_{\rm in})=\frac{1}{\pi}\left(2r_{\rm in}^{T}r_{\rm
in}-1\right)\mbox{exp}\left(-r_{\rm in}^{T}r_{\rm in}\right),
\end{eqnarray}
where $r_{\rm in}=\left(x_{\rm in},p_{\rm in}\right)^{T}$ is the
radius vector in phase space. In the origin the function attains
minimum possible negative value allowed by quantum mechanics equal
to $W_{\rm in}(0)=-1/\pi\doteq-0.3181$, where here $0$ stands for
a zero $2\times1$ vector.

We consider a standard CV teleportation protocol
\cite{Braunstein_98,Furusawa_98} in the nonunity gain regime
\cite{Bowen_03}. An input mode characterized by the quadrature
operators $x_{\rm in},p_{\rm in}$ satisfying the canonical commutation rules
$\left[x_{\rm in},p_{\rm in}\right]=i$ prepared in Fock state $|1\rangle$ is
teleported by Alice ($A$) to Bob ($B$). Initially, Alice and Bob
hold modes $A$ and $B$, respectively, described by the quadratures
$x_{i},p_{i}$, $i=A,B$, in a pure two-mode squeezed vacuum state
with squeezed Einstein-Podolsky-Rosen variances
$\langle\left[\Delta\left(x_{A}-x_{B}\right)\right]^2\rangle=\langle\left[\Delta\left(p_{A}+p_{B}\right)\right]^2\rangle=e^{-2r}$,
where $r$ is the squeezing parameter. The state can be prepared by
mixing of two pure squeezed states with squeezed variances $V_{\rm
sq}=\langle\left(\Delta p_{A}\right)^2\rangle=\langle\left(\Delta
x_{B}\right)^2\rangle=e^{-2r}/2$ on a balanced beam splitter.
Next, Alice superimposes the input mode with mode $A$ of the
shared entangled state on an unbalanced beam splitter with
reflectivity $\sqrt{R}$ and transmissivity $\sqrt{T}$ ($R+T=1$)
and measures the quadratures $x_{u}=\sqrt{R}x_{\rm
in}-\sqrt{T}x_{A}$ and $p_{v}=\sqrt{T}p_{\rm in}+\sqrt{R}p_{A}$ at
the outputs of the beam splitter. She then sends the measurement
outcomes $\bar{x}_{u},\bar{p}_{v}$ via classical channel to Bob
who displaces his mode $B$ as $x_{B}\rightarrow
x_{B}+g_{x}\bar{x}_{u}$, $p_{B}\rightarrow
p_{B}+g_{p}\bar{p}_{v}$, where $g_{x},g_{p}$ are electronic gains,
thereby partially recreating the input state on mode $B$.

\begin{figure}
\centerline{\psfig{width=9.0cm,angle=0,file=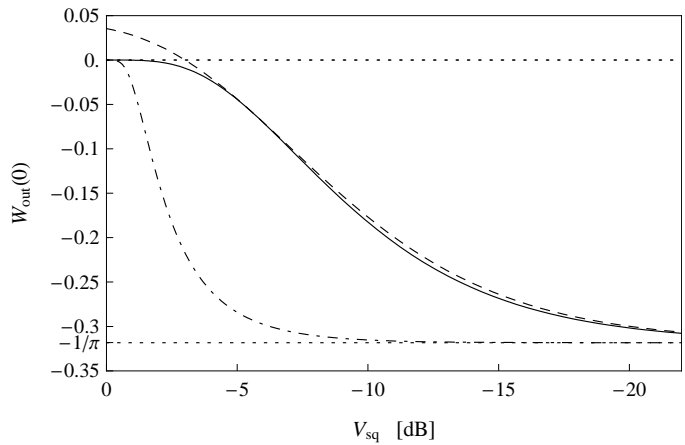}}
\caption{Output Wigner function in the origin versus the squeezed
variance $V_{\rm sq}$ for optimal nonunity gain teleportation
(solid curve), unity gain teleportation (dashed curve) and optimal
conditional teleportation with $K=0.3$ (dash-dotted curve) of a
single-photon Fock state. See text for details.}\label{figure1}
\end{figure}

From the mathematical point of view the nonunity gain
teleportation belongs to the class of single-mode trace-preserving
Gaussian completely positive maps \cite{Lindblad_00}. On the level
of Wigner functions such a map transforms the Wigner function of
the input state $W_{\rm in}(r_{\rm in})$ according to the integral
formula \cite{Fiurasek_02}:
\begin{eqnarray}\label{WCPout}
W_{\rm out}(r_{\rm
out})=2\pi\int_{-\infty}^{+\infty}W_{\chi}(r_{\rm in},r_{\rm
out})W_{\rm in}(\Lambda r_{\rm in})dr_{\rm in},
\end{eqnarray}
where $r_{\rm out}=\left(x_{\rm out},p_{\rm out}\right)^{T}$,
$\Lambda=\mbox{diag}(1,-1)$ and $W_{\chi}$ is the following
two-mode Gaussian kernel:
\begin{eqnarray}\label{Wchi}
W_{\chi}(r_{\rm in},r_{\rm
out})=\frac{1}{2\pi^2\sqrt{\mbox{det}Q}}\mbox{exp}\left(-\Delta
r^{T}Q^{-1}\Delta r\right),
\end{eqnarray}
where $\Delta r=r_{\rm out}-S\Lambda r_{\rm in}$, $Q$ is a real
symmetric positive semidefinite $2\times 2$ matrix, $S$ is a real
$2\times 2$ matrix and the matrices satisfy the inequality
$Q+iJ-iSJS^{T}\geq0$, where \small $J=\left(\begin{array}{cc}
0 & 1 \\
-1 & 0\\
\end{array}\right)$\normalsize. For nonunity gain teleportation we have, in particular,
\begin{eqnarray}\label{S}
S=\left(\begin{array}{cc}
g_{x}\sqrt{R} & 0 \\
0 & g_{p}\sqrt{T}\\
\end{array}\right)
\end{eqnarray}
and $Q=\mbox{diag}(Q_{x},Q_{p})$, where
\begin{eqnarray}\label{Q}
Q_{x}&=&\cosh(2r)+g_{x}^{2}T\cosh(2r)-2g_{x}\sqrt{T}\sinh(2r),\nonumber\\
Q_{p}&=&\cosh(2r)+g_{p}^{2}R\cosh(2r)-2g_{p}\sqrt{R}\sinh(2r).\nonumber\\
\end{eqnarray}
Substituting from Eqs.~(\ref{S}) and (\ref{Q}) into
Eq.~(\ref{Wchi}) and calculating the integral in
Eq.~(\ref{WCPout}) for the input Wigner function given by
Eq.~(\ref{Win}) we find the output Wigner function in the origin
in the form:
\begin{eqnarray}\label{Wout}
W_{\rm
out}(0)=\frac{\mbox{det}Q-\left(\mbox{det}S\right)^2}{\pi\left[\mbox{det}\left(SS^{T}+Q\right)\right]^{\frac{3}{2}}}.
\end{eqnarray}
From practical point of view it is important to know what is the
largest negative value of the Wigner function that can be obtained
at the output of the teleportation for a given level of shared
entanglement. This requires minimization of the function
(\ref{Wout}) for a fixed $r$ over three variables $g_{x},g_{p}$
and $T$ which can be barely done analytically. Numerical
minimization, however, indicates that as one would expect optimal
performance of the teleportation is achieved if the beam splitter
is balanced, i.e., $\sqrt{R}=\sqrt{T}=1/\sqrt{2}$, and if the
teleportation adds noise symmetrically into position and momentum
quadrature, i.e. $g_{x}=g_{p}=g$. Under these assumptions and
introducing the normalized gain $G=g/\sqrt{2}$ we can express the
Wigner function in the origin as
\begin{eqnarray}\label{WoutG}
W_{\rm out}(0)=\frac{\alpha
(G)-G^2}{\pi\left[\alpha(G)+G^2\right]^{2}},
\end{eqnarray}
where $\alpha(G)=\cosh(2r)(1+G^{2})-2G\sinh(2r)$. Solving the
extremal equation $dW_{\rm out}(0)/dG=0$ with respect to $G$ one
finds that optimal gain can be found as a root of the following
third-order polynomial
\begin{eqnarray}\label{Gpolynomial}
G^{3}+aG^{2}+bG+c=0,
\end{eqnarray}
where
\begin{eqnarray}\label{abc}
a&=&-3\coth(r),\quad b=2+\coth^{2}(2r)+3\frac{\cosh(2r)}{\sinh^{2}(2r)},\nonumber\\
c&=&-\coth(2r).
\end{eqnarray}
The polynomial has three real roots of the form
$G_{1,2,3}=y_{1,2,3}+\coth(r)$, where
\begin{eqnarray}\label{roots}
y_{1}=2\sqrt{-\frac{p}{3}}\cos\left(\frac{\phi}{3}\right),\:
y_{2,3}=-2\sqrt{-\frac{p}{3}}\cos\left(\frac{\phi\pm\pi}{3}\right),\nonumber\\
\end{eqnarray}
where $\cos\phi=-(q/2)\sqrt{-27/p^3}$ and
\begin{equation}\label{pq}
p=b-\frac{a^2}{3},\quad q=c-\frac{ab}{3}+\frac{2a^3}{27}.
\end{equation}
Substituting the roots $G_{1,2,3}$ back into the right hand side
of Eq.~(\ref{WoutG}) and plotting the obtained function in
dependence of the squeezed variance $V_{\rm sq}$ one finds the
optimal gain $G_{\rm opt}$ minimizing the output Wigner function
in the origin to be $G_{\rm opt}=G_{2}$. The output Wigner
function in the origin in dependence of the squeezed variance
$V_{\rm sq}$ for optimal nonunity gain teleportation is depicted
by solid curve in Fig.~\ref{figure1}.
The figure reveals that the Wigner function in the origin is
monotonously decreasing function of the squeezing approaching the
minimum value of $W_{\rm in}(0)=-1/\pi\doteq-0.3181$ in the limit
of infinitely large squeezing. The figure further shows that
optimal nonunity gain teleportation transfers successfully the
negative values of the Wigner function for arbitrarily small
nonzero squeezing $r>0$. The latter finding should be contrasted
with the unity gain regime that is recovered for
$\sqrt{R}=\sqrt{T}=1/\sqrt{2}$ and $g_{x}=g_{p}=\sqrt{2}$. Then,
equations (\ref{S}) and (\ref{Q}) give $S=\openone$,
$Q=2e^{-2r}\openone$ that leads using Eq.~(\ref{Wout}) to the
output Wigner function in the origin in the form
\begin{eqnarray}\label{Woutunitygain}
\tilde{W}_{\rm
out}(0)=\frac{2e^{-2r}-1}{\pi\left(2e^{-2r}+1\right)^{2}}.
\end{eqnarray}
Hence it immediately follows that in the unity gain regime the
output Wigner function in the origin is negative only if
$e^{-2r}<1/2$, i.e., if the squeezing is larger than -3 dB (see
also dashed curve in Fig.~\ref{figure1}) which corresponds to the
fidelity benchmark $F=2/3$ \cite{Ban_Caves_04}. Thus while
nonunity gain teleportation allows teleportation of a negative
Wigner function of the Fock state $|1\rangle$ for an arbitrarily
small squeezing unity gain teleportation requires more than -3 dB
squeezing to accomplish this task. For comparison we mention
explicitly the value of the output Wigner function in the origin
for nonunity and unity gain regimes for several values of
squeezing. For -3 dB squeezing the optimal nonunity gain
teleportation gives $W_{\rm out}(0)\doteq -0.0091$ while the unity
gain teleportation yields $\tilde{W}_{\rm out}(0)\doteq 0.0002$,
for -5 dB we get $W_{\rm out}(0)\doteq -0.0442$ and
$\tilde{W}_{\rm out}(0)\doteq -0.0439$, for -7 dB we get $W_{\rm
out}(0)\doteq -0.0993$ and $\tilde{W}_{\rm out}(0)\doteq -0.0977$,
and for -10 dB we get $W_{\rm out}(0)\doteq -0.1826$ and
$\tilde{W}_{\rm out}(0)\doteq -0.1768$.

\begin{figure}
\centerline{\psfig{width=8.5cm,angle=0,file=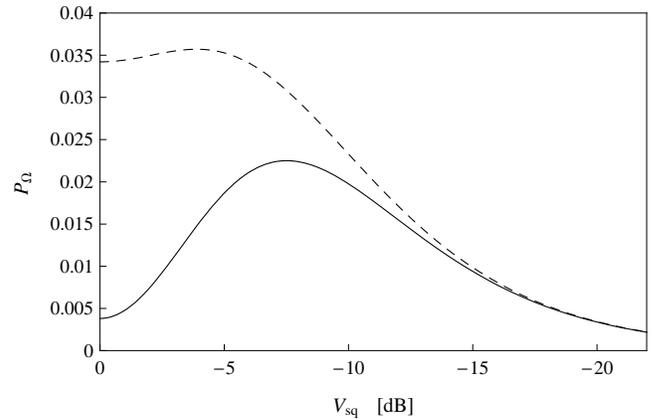}}
\caption{Success probability $P_{\Omega}$ versus the squeezed
variance $V_{\rm sq}$ for conditional teleportation of the
single-photon Fock state (solid curve) and the state $\rho_{\eta}$
with $\eta=0.6304$ (dashed curve) for $K=0.3$. See text for
details.}\label{figure2}
\end{figure}

Summarizing the obtained results we see that for the single-photon
Fock state at the input of the teleportation we can get a state
with a negative Wigner function in the origin at the output of the
teleportation for arbitrarily small squeezed variance $V_{\rm sq}$
provided that the gain of the teleportation is adjusted suitably.
Achievement of a reasonably high negativity not less than an order
of the magnitude smaller than the negativity at the input,
however, requires squeezed variances larger than -5 dB.
Substantially larger negative values for lower squeezed variances
are obtained by using the {\it conditional} teleportation where we
accept the output state only when the outcome of Alice's
measurement $\beta\equiv (\bar{x}_{u}+i\bar{p}_{v})/\sqrt{2}$
falls inside a circle centered in the origin with radius $K$,
i.e., falls into the set $\Omega=\left\{\beta,|\beta|\leq K,
K>0\right\}$. If a measurement outcome $\beta$ was detected then
the unnormalized output state is \cite{Ide_02_ng}
\begin{eqnarray}\label{psibeta}
|\psi(\beta)\rangle&=&\sqrt{1-\lambda^{2}}e^{-(1-\lambda^2)\frac{|\beta|^2}{2}}D\left[(G-\lambda)\beta\right]\nonumber\\
&&\times\left[(1-\lambda^2)\beta^{\ast}|0\rangle+\lambda|1\rangle\right],
\end{eqnarray}
where $\lambda=\tanh r$ and $D(\alpha)=\mbox{exp}(\alpha
a^{\dag}-\alpha^{\ast}a)$ is the displacement operator. The
probability of finding the outcome in the set $\Omega$ then reads
\begin{eqnarray}\label{POmega}
P_{\Omega}&=&\frac{1}{\pi}\int_{\Omega}\langle\psi(\beta)|\psi(\beta)\rangle d^{2}\beta\nonumber\\
&=&1-\left[1+(1-\lambda^2)^{2}K^{2}\right]e^{-(1-\lambda^2)K^2}
\end{eqnarray}
and the normalized density matrix of the output state is
\begin{eqnarray}\label{rhoOmega}
\rho_{\Omega}&=&\frac{1}{\pi
P_{\Omega}}\int_{\Omega}|\psi(\beta)\rangle\langle\psi(\beta)|d^{2}\beta.
\end{eqnarray}
The Wigner function in the origin of the state is then easy to
calculate as the expectation value
$W_{\Omega}(0)=\mbox{Tr}\left[\rho_{\Omega}(-1)^{n}\right]/\pi$
\cite{Royer_77} of the parity operator $(-1)^{n}$. Substituting
into the latter formula from Eqs.~(\ref{psibeta}), (\ref{POmega})
and (\ref{rhoOmega}) and performing the integration over $\beta$
we arrive at the following output Wigner function in the origin
\begin{eqnarray}\label{Woutprob}
W_{\Omega}(0)&=&\frac{(1-\lambda^2)}{\pi
P_{\Omega}}\left\{-\frac{\lambda^{2}}{a}\left(1-e^{-aK^{2}}\right)\right.\nonumber\\
&&\left.+\frac{(\lambda^2-2G\lambda+1)^{2}}{a^2}\left[1-\left(1+aK^{2}\right)e^{-aK^{2}}\right]\right\},\nonumber\\
\end{eqnarray}
where $a=(1-\lambda^2)+2(G-\lambda)^2$. As post-selection interval
$K$ vanishes, the role of optimized displacement becomes
negligible and the Wigner function in the origin approaches
original value $W_{\Omega}(0)=-1/\pi$ of the single photon state,
irrespectively to the squeezing used to produce the shared
entangled state. It corresponds to previously obtained result for
the fidelity of teleportation \cite{Hofmann_01,Ide_02_ng}. We
performed numerical optimization of the gain $G$ and depicted the
Wigner function in the origin (\ref{Woutprob}) by the dash-dotted
curve in Fig.~\ref{figure1} for $K=0.3$. The corresponding success
probability $P_{\Omega}$ is depicted by the solid curve in
Fig.~\ref{figure2}.

The figure shows that conditional teleportation substantially
outperforms the optimal unconditional teleportation, of course, at
the expense of the probabilistic nature of the protocol. For
example, conditional teleportation with $K=0.3$ gives for -3 dB
squeezing $P_{\Omega}=0.0112$ and $W_{\Omega}(0)\doteq -0.2174$,
for -5 dB we get $P_{\Omega}=0.0187$ and $W_{\Omega}(0)\doteq
-0.284$, for -7 dB we get $P_{\Omega}=0.0223$ and
$W_{\Omega}(0)\doteq -0.3056$ and for -10 dB we get
$P_{\Omega}=0.0198$ and $W_{\Omega}(0)\doteq -0.3152$. The
obtained values indicate that conditional teleportation allows
to achieve high negative values of the Wigner function in the
origin even for moderate levels of squeezing approximately equal
to -3 dB at a cost of roughly 1.1\% probability of success.

\begin{figure}
\centerline{\psfig{width=8.5cm,angle=0,file=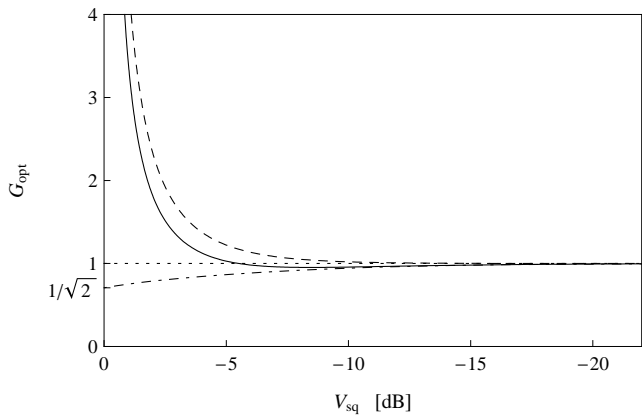}}
\caption{Optimal normalized gain $G_{\rm opt}$ (solid curve)
versus the squeezed variance $V_{\rm sq}$ for teleportation of a
single-photon Fock state. The dashed curve corresponds to the gain
of teleportation minimizing the output added noise \cite{Ralph_00}
and the dash-dotted curve corresponds to the gain of teleportation
maximizing the average teleportation fidelity \cite{Ide_02}. See
text for details.}\label{figure3}
\end{figure}

In order to get a deeper insight into the performance of the
teleportation that is optimal for teleportation of the Wigner
function in the origin for the Fock state $|1\rangle$ we display
by a solid curve in Fig.~\ref{figure3} the optimal gain $G_{\rm
opt}$ as a function of the squeezed variance $V_{\rm sq}$.
We see from the figure that for squeezing less than -5.52 dB
optimal teleportation works as a phase-insensitive amplifier while
for larger squeezing it is a weak attenuator approaching the unity
gain regime in the limit of infinitely large squeezing. It is of
interest to compare the optimal gain with the gain of
teleportation that is optimal in the sense that it adds for a
given added noise into Alice's measurement outcomes the least
possible noise into the output state \cite{Ralph_00}. In this
protocol the optimal gain depends on the squeezing of the shared
two-mode squeezed vacuum state as $G_{\rm opt}'=\coth(2r)$ and it
is depicted by the dashed curve in Fig.~\ref{figure3}. It is
clearly visible from the figure that teleportation adding minimum
noise is not optimal for teleportation of a Wigner function in the
origin of the Fock state $|1\rangle$. While the first
teleportation is a phase-insensitive amplifier for all levels of
squeezing the latter one acts like a phase-insensitive attenuator
for squeezing larger than -5.52 dB. We should also stress here
that our teleportation protocol that is optimal from the point of
view of the output Wigner function in the origin differs from the
optimal teleportation of the single-photon Fock state maximizing
the average teleportation fidelity that was investigated in
\cite{Ide_02}. In the latter protocol the optimal normalized gain
depicted by the dash-dotted curve in Fig.~\ref{figure3} always
lies between $1/\sqrt{2}$ and $1$ and therefore the teleportation
maximizing the average teleportation fidelity realizes a
phase-insensitive attenuator for all levels of squeezing.

\begin{figure}[h]
\centerline{\psfig{width=7.5cm,angle=0,file=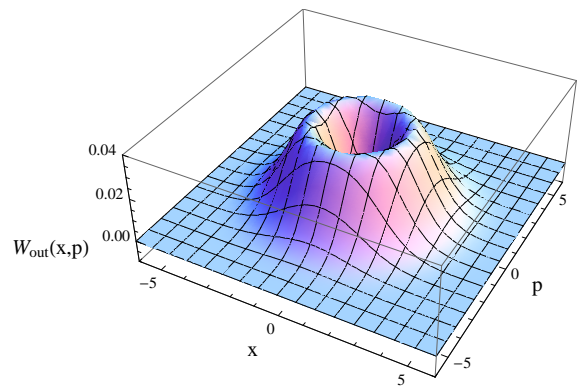}}
\caption{Wigner function of the output state for the optimal
nonunity gain teleportation for $e^{2t}=2$, $V_{\rm sq}=-3$ dB and
transmissivity $\sqrt{T}=1/\sqrt{3}$. The Wigner function in the
origin attains the negative value of $W_{\rm out}^{(\rm
sq)}(0)\doteq -9.10^{-3}$. See text for details.}\label{figure4}
\end{figure}

\begin{figure}[h]
\centerline{\psfig{width=7.5cm,angle=0,file=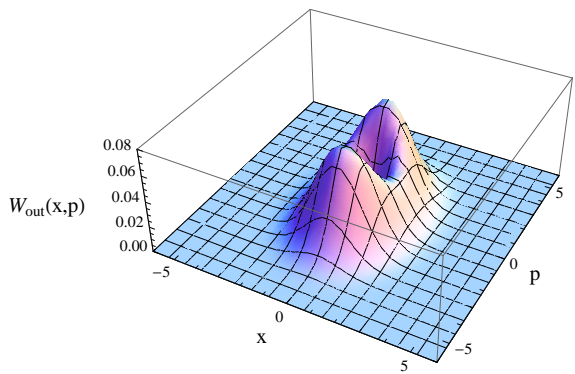}}
\caption{Wigner function of the output state for the unity gain
teleportation for $e^{2t}=2$ and $V_{\rm sq}=-3$ dB. The Wigner
function is equal to zero in the origin.}\label{figure5}
\end{figure}

Up to now we considered teleportation of the Fock state
$|1\rangle$. In practice, the states with a negative Wigner
function are prepared by a single-photon subtraction from a
squeezed state \cite{Dakna_97}. The subtraction is implemented by
mixing of a squeezed state squeezed in the position quadrature
$x_{\rm in}$ with variance $\langle\left(\Delta x_{\rm
in}\right)^2\rangle=e^{-2s}/2$ on a beam splitter with amplitude
transmissivity $\sqrt{\tau}$ followed by projection of one of its
outputs on the Fock state $|1\rangle$. As a result we obtain the
squeezed single-photon Fock state $S(t)|1\rangle$, where
$S(t)=\mbox{exp}[(t/2)(a^2-a^{{\dag}^2})]$ is the squeezing
operator and $t$ is the squeezing parameter satisfying $\tanh
t=\tau\tanh s$. The state has the Wigner function in the origin of
the form:
\begin{eqnarray}\label{Wsqin}
W_{\rm in}^{(\rm sq)}(r_{\rm in})=\frac{1}{\pi}\left(2r_{\rm
in}^{T}\gamma^{-1}r_{\rm in}-1\right)\mbox{exp}\left(-r_{\rm
in}^{T}\gamma^{-1}r_{\rm in}\right),
\end{eqnarray}
where $\gamma=\mbox{diag}(e^{-2t},e^{2t})$. Substituting the
Wigner function into the formula (\ref{WCPout}) and carrying out
the integration we arrive at the following output Wigner function
\begin{eqnarray}\label{Wsqout}
W_{\rm out}^{(\rm sq)}(r_{\rm out})&=&\left[2r_{\rm
out}^{T}Zr_{\rm
out}+\frac{\mbox{det}Q-\left(\mbox{det}S\right)^2}{\mbox{det}\tilde{\gamma}}\right]\nonumber\\
&&\times\frac{\mbox{exp}\left(-r_{\rm
out}^{T}\tilde{\gamma}^{-1}r_{\rm
out}\right)}{\pi\sqrt{\mbox{det}\tilde{\gamma}}},
\end{eqnarray}
where $Z=\tilde{\gamma}^{-1}S\gamma S^{\rm T}\tilde{\gamma}^{-1}$,
$\tilde{\gamma}=S\gamma S^{\rm T}+Q$ and the matrices $S$ and $Q$
are given in Eqs.~(\ref{S}) and (\ref{Q}). In the origin the
output Wigner function is equal to
\begin{eqnarray}\label{Wsqueezedout}
W_{\rm out}^{(\rm
sq)}(0)=\frac{\mbox{det}Q-\left(\mbox{det}S\right)^2}{\pi\left[\mbox{det}\left(S\gamma
S^{T}+Q\right)\right]^{\frac{3}{2}}}.
\end{eqnarray}
Teleportation of the squeezed single-photon Fock state can be
easily transformed into the optimal teleportation of the
single-photon Fock state. Obviously, it is sufficient if the
teleportation simply compensates the squeezing represented by the
covariance matrix $\gamma$ and simultaneously its overall
normalized gain is equal to $G_{\rm opt}$. Indeed, returning back
to the more general protocol with transmissivity $\sqrt{T}$ and
gains $g_{x,p}$ and setting the gains as $g_{x}=e^{t}G_{\rm
opt}/\sqrt{R}$, $g_{p}=e^{-t}G_{\rm opt}/\sqrt{T}$ and the
transmissivity such that $T/R=e^{-2t}$ one finds that $S=G_{\rm
opt}\mbox{diag}(e^{t},e^{-t})$, $S\gamma S^{T}=G_{\rm
opt}^{2}\openone$ and $Q=\alpha(G_{\rm opt})\openone$, where
$\alpha(G)$ is given below Eq.~(\ref{WoutG}). Substitution of the
latter expressions of $S\gamma S^{T}$ and $Q$ into
Eq.~(\ref{Wsqueezedout}) leads finally to the minimal output
Wigner function in the origin for Fock state $|1\rangle$. In order
to illustrate the marked difference between the value of the
output Wigner function in the origin as well as its shape for the
optimal nonunity gain teleportation of the squeezed single-photon
state and unity gain teleportation of the state we plot the entire
output Wigner functions for the two scenarios in
Figs.~\ref{figure4} and \ref{figure5}.

\section{Teleportation of an attenuated single-photon Fock
state}\label{sec_2}

\begin{figure}
\centerline{\psfig{width=8.0cm,angle=0,file=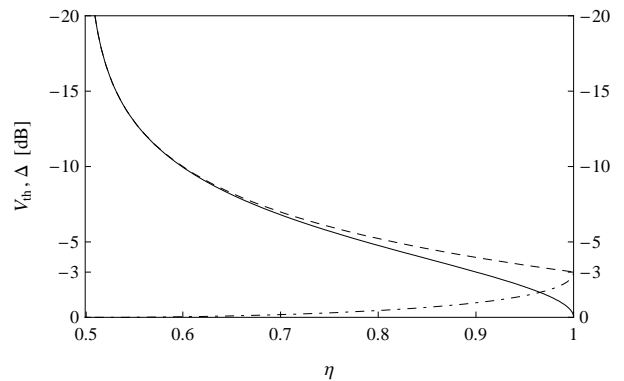}}
\caption{Threshold squeezed variances $V_{\rm th}^{(G)}$ (solid
curve), $V_{\rm th}^{(1)}$ (dashed curve) and their difference
$\Delta=V_{\rm th}^{(1)}-V_{\rm th}^{(G)}$ (dash-dotted curve)
versus the probability $\eta$. See text for
details.}\label{figure6}
\end{figure}

\begin{figure}
\centerline{\psfig{width=8.5cm,angle=0,file=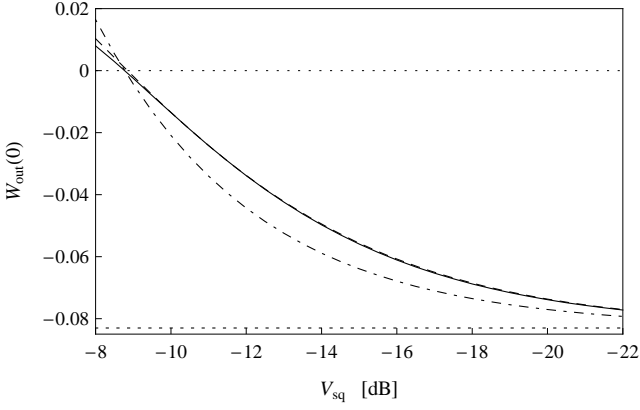}}
\caption{Output Wigner function in the origin versus the squeezed
variance $V_{\rm sq}$ for optimal nonunity gain teleportation
(solid curve), unity gain teleportation (dashed curve) and optimal
conditional teleportation with $K=0.3$ (dash-dotted curve) for
the input state $\eta|1\rangle\langle 1|+(1-\eta)|0\rangle\langle
0|$ with $\eta=0.6304$. Bottom dotted curve corresponds to the
input Wigner function in the origin $W^{(\eta)}_{\rm
in}(0)\doteq-0.083$. See text for details.}\label{figure7}
\end{figure}

\begin{figure}
\centerline{\psfig{width=8.0cm,angle=0,file=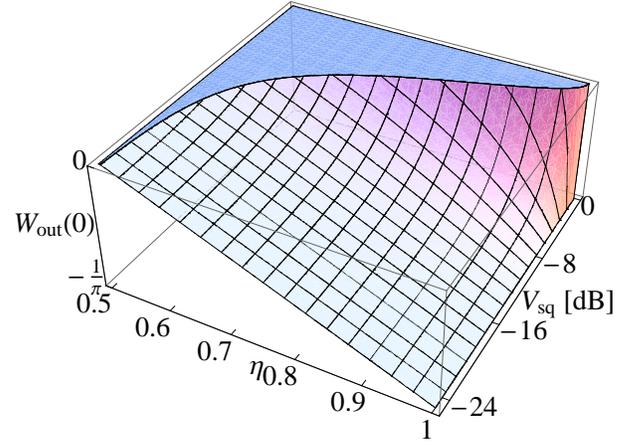}}
\caption{Asymptotic Wigner function in the origin (\ref{WOmegaK0})
for the attenuated single-photon state $\rho_{\eta}$ after the
conditional teleportation in dependence on the probability $\eta$
and the squeezed variance $V_{\rm sq}$.}\label{figure8}
\end{figure}

Quantum states with a negative Wigner function prepared currently
in a laboratory have a substantially reduced negativity in
comparison with Fock state $|1\rangle$ and they are mixed. From an
experimental point of view it is therefore imperative to know the
bounds one has to surpass in order to successfully teleport mixed
states with a negative Wigner function. In an experiment the main
source of mixedness are losses that can be in the case of the Fock
state $|1\rangle$ most simply modelled by a purely lossy channel
that transmits the state with probability $\eta$ and replaces it
by the vacuum state with probability $1-\eta$. At the output of
the channel we get the mixed state
\begin{equation}\label{rhoeta}
\rho_{\eta}=\eta|1\rangle\langle 1|+(1-\eta)|0\rangle\langle 0|
\end{equation}
with Wigner function in the origin equal to $W^{(\eta)}_{\rm
in}(0)=(1-2\eta)/\pi$ that is negative if $\eta>1/2$. Making use
of the formula (\ref{WCPout}) where we set $S=G\openone$ and
$Q=\alpha(G)\openone$ we arrive at the output Wigner function in
the origin in the form:
\begin{eqnarray}\label{Weta}
W^{(\eta)}_{\rm out}(0)=\frac{1}{\pi}\left\{\eta\frac{\alpha
(G)-G^2}{\left[\alpha(G)+G^2\right]^{2}}+\frac{1-\eta}{\alpha(G)+G^2}\right\}.
\end{eqnarray}
The formula allows us to calculate for a given probability $\eta$
the threshold value of the squeezing above which the output state
has a negative Wigner function in the origin. From the condition
$W^{(\eta)}_{\rm out}(0)<0$ we therefore obtain after some algebra
that the Wigner function in the origin (\ref{Weta}) is negative if
the squeezing parameter $r$ satisfies $r>r_{\rm
th}^{(G)}=\operatorname{arctanh}\sqrt{(1-\eta)/\eta}$. By setting
$G=1$ in the formula (\ref{Weta}) and repeating the above
calculation one finds, on the other hand, that the output Wigner
function in the origin for unity gain teleportation is negative if
the squeezing parameter $r$ satisfies $r>r_{\rm
th}^{(1)}=\ln\sqrt{2/(2\eta-1)}$. In Fig.~\ref{figure6} we plot
the dependence of the threshold squeezed variances $V_{\rm
th}^{(G)}=e^{-2r_{\rm th}^{(G)}}/2$ and $V_{\rm
th}^{(1)}=e^{-2r_{\rm th}^{(1)}}/2$ on the probability $\eta$.
It is apparent from the figure that starting from $\eta=1$ the
squeezing costs increase slowly with decreasing probability $\eta$
up to $\eta\approx0.6$. For probabilities less than approximately
$0.6$ that correspond to the negative values of the Wigner
function already demonstrated experimentally the squeezing costs
increase dramatically as $\eta$ approaches $\eta=0.5$. As an
illustrative example consider the state $\rho_{\eta}$ with
$\eta=0.6304$ corresponding to $W^{(\eta)}_{\rm
in}(0)\doteq-0.083$ which was recently achieved experimentally
\cite{Wakui_07}. In order to have for the state the output Wigner
function in the origin negative we need the squeezing parameter
$r>r_{\rm th}^{(G)}=1.0098$ corresponding to the squeezed variance
$V_{\rm sq}$ larger than -8.77 dB.  Further, the threshold
squeezing is apparently lower for the optimal nonunity gain
teleportation than for the unity gain teleportation and the
difference increases with increasing probability $\eta$ up to -3
dB for $\eta=1$. In Fig.~\ref{figure7} we plot the output Wigner
function in the origin for the state with $\eta=0.6304$ versus the
squeezed variance $V_{\rm sq}$.
The Fig.~\ref{figure7} reveals relatively steep decrease of the
Wigner function in the origin with increasing squeezing for
squeezed variance up to $V_{\rm sq}\approx -14$ dB. For larger
squeezing a saturation effect occurs when a small decrease of the
value of the Wigner function in the origin requires a large
increase of the squeezing. The figure also illustrates that
observation of a reasonably large negativity of the Wigner
function at the output of teleportation will require the highest
squeezing levels ever achieved. For example, a squeezed variance
$V_{\rm sq}=-10$ dB that was recently observed experimentally
\cite{Vahlbruch_08} would yield the output Wigner function in the
origin for optimal nonunity gain teleportation equal to
$W^{(\eta)}_{\rm out}(0)\doteq-0.0135$. Further improvement can be
reached again by using the conditional teleportation with
optimized gain. For the state $\rho_{\eta}$ at the input we get
the output Wigner function in the origin in the form:
\begin{eqnarray}\label{Wp}
W_{\Omega}^{(\eta)}(0)=\eta
W_{\Omega}^{(1)}(0)+(1-\eta)W_{\Omega}^{(0)}(0),
\end{eqnarray}
where
$W_{\Omega}^{(1)}(0)=(P_{\Omega}/P_{\Omega}^{(\eta)})W_{\Omega}(0)$
and
\begin{eqnarray}\label{W0}
W_{\Omega}^{(0)}(0)=\frac{(1-\lambda^{2})}{\pi
P_{\Omega}^{(\eta)}a}\left(1-e^{-aK^2}\right),
\end{eqnarray}
where
\begin{eqnarray}\label{POmegaeta}
P_{\Omega}^{(\eta)}&=&1-\left[1+\eta(1-\lambda^2)^{2}K^{2}\right]e^{-(1-\lambda^2)K^2}
\end{eqnarray}
is the success probability, where $P_{\Omega}$ is defined in
Eq.~(\ref{POmega}) and $a$ is defined below Eq.~(\ref{Woutprob}).
As post-selection interval $K$ vanishes, the Wigner function in origin approaches the lowest value
\begin{equation}\label{WOmegaK0}
W_{\Omega,K=0}^{(\eta)}(0)=\frac{1}{\pi}\frac{1-\eta-\eta\lambda^2}{1-\eta+\eta\lambda^2},
\end{equation}
which can be achieved by conditional Gaussian teleportation of the
attenuated single-photon state, at the cost of success rate. The
threshold to preserve the negativity of the Wigner function is
clearly the same as for the unconditional teleportation, i.e.,
$r_{\rm th}^{\rm(cond)}=r_{\rm
th}^{(G)}=\operatorname{arctanh}\sqrt{(1-\eta)/\eta}$. For large
squeezing levels the Wigner function (\ref{WOmegaK0}) can be
expanded in the parameter $\lambda$ around the point 1 as
\begin{equation}
W_{\Omega,K=0}^{(\eta)}(0)\approx\frac{1}{\pi}\left[1-2\eta+4\eta\left(1-\eta\right)\left(1-\lambda\right)\right].
\end{equation}
Hence it follows that comparing to the ideal single-photon Fock
state, the value of the Wigner function in the origin of the state
after the teleportation approaches the initial value $W_{\rm
in}^{(\eta)}(0)=\left(1-2\eta\right)/\pi$ only in the limit of
infinitely large squeezing used to prepare the entangled state,
i.e., for $\lambda\rightarrow 1$. How much squeezing is required
is clearly visible from Fig.~\ref{figure8}. This is a substantial
difference from the idealized single-photon state for which the
conditional teleportation approaches unit fidelity for an
arbitrary small squeezing used for the production of the shared
entangled state.

To compare negativity of the output Wigner function for the
attenuated single-photon state for conditional and unconditional
teleportation we plotted the output Wigner function in the origin
(\ref{Wp}) for $K=0.3$ by the dash-dotted curve in
Fig.~\ref{figure7} and the corresponding success probability
(\ref{POmegaeta}) by the dashed curve in Fig.~\ref{figure2}. The
gain of the conditional teleportation is optimized numerically and
it differs from the optimal gain of the unconditional
teleportation on at most third decimal place. It is apparent from
the figure that conditional teleportation pronouncedly outperforms
the optimal unconditional nonunity gain teleportation. For
instance, for the state $\rho_{\eta}$ with $\eta=0.6304$ and
squeezed variance $V_{\rm sq}=-10$ dB we obtain
$W_{\Omega}^{(\eta)}(0)\doteq-0.0209$ and the corresponding
success probability is $P_{\Omega}^{(\eta)}\doteq0.0233$ in
comparison with $W^{(\eta)}_{\rm out}(0)\doteq-0.0135$ that is
obtained for unity gain regime. The figure also reveals that for
both the teleportations the output Wigner function in the origin
becomes negative for the same value of the threshold squeezed
variance equal to -8.77 dB which is in full accordance with our
finding of impossibility to reduce the threshold squeezing by
resorting to a conditional protocol. It is also worth mentioning
that the optimal teleportation is again an amplifier for squeezed
variance less than -10.84 dB and then changes to a weak attenuator
for larger squeezing that finally approaches unity gain regime in
the limit of infinitely large squeezing. In comparison with the
case of the pure Fock state $|1\rangle$ depicted in
Fig.~\ref{figure1} the advantage of the nonunity gain regime for
the mixed state $\rho_{\eta}$ wipes out.
\section{Conditional teleportation with noise excess} \label{sec_3}

As we already said in the introduction quantum teleportation is a
basic building block for quantum computation and long-distance
quantum communication. In quantum computation unconditional
quantum teleportation can be used for implementation of a
deterministic gate on an arbitrary quantum state. Therefore,
deterministic transmission of a negative Wigner function by
unconditional teleportation is a necessary prerequisite for
successful gate operation. Previous analysis indicates that even
for an ideal input Fock state $|1\rangle$ the state at the output
of unconditional teleportation will have for realistic squeezing
levels substantially reduced negative value of the Wigner function
in the origin. In addition, it will be mixed so that successful
transmission of the negativity of the Wigner function of this
state through the next gate will require even larger squeezing
owing to the existence of a strict bound on the minimum squeezing
needed to teleport a negative Wigner function of a mixed state.

On the contrary, for quantum communication purposes it suffices to
implement just conditional teleportation which will only reduce
the success rate of anyway probabilistic communication protocol.
Since the conditional teleportation gives for currently achievable
levels of squeezing better results than the unconditional one and
already finds application in quantum communication in this section
we will restrict ourself to the analysis of conditional
teleportation. In previous sections we assumed an ideal case of
pure shared entanglement produced by mixing of two pure squeezed
states. Here we will do another step towards a more realistic
scenario by considering impure squeezed states with noise excess
in the anti-squeezed quadrature. We will show that conditional
teleportation of the negative Wigner function of the Fock state
$|1\rangle$ is tolerable to the realistic values of the noise
excess.

Let us therefore consider the squeezed states of modes $A$ and $B$
to be momentum and position squeezed states with squeezed
quadratures $V_{\rm sq}=\langle\left(\Delta
p_{A}\right)^2\rangle=\langle\left(\Delta x_{B}\right)^2\rangle$
and the anti-squeezed quadratures $V_{\rm an}=\langle\left(\Delta
x_{A}\right)^2\rangle=\langle\left(\Delta p_{B}\right)^2\rangle$.
Let us further denote the input state $\rho_{\rm in}$, the shared
entangled state $\rho_{AB}$ and by $\Pi_{{\rm in} A}(\beta)$ the
projector onto the Bell state $|\beta\rangle_{\rm in
A}=\sum_{n=0}^{\infty}D_{\rm in}(\beta)|n\rangle_{\rm
in}|n\rangle_{A}$ \cite{Hofmann_00}, where $D(\beta)$ is the
displacement operator defined below Eq.~(\ref{psibeta}). The state
at the output of the teleportation conditioned on the measurement
outcome $\beta= (\bar{x}_{u}+i\bar{p}_{v})/\sqrt{2}$ and displaced
according to the measurement outcome with normalized gain $G$ by
Bob then reads
\begin{eqnarray}\label{rhobeta}
\tilde{\rho}_{\rm B}(\beta)=D_{B}(G\beta)\mbox{Tr}_{{\rm
in}A}\left[\rho_{\rm in}\otimes\rho_{AB}\Pi_{{\rm in}
A}(\beta)\right]D_{B}^{\dag}(G\beta).\nonumber\\
\end{eqnarray}
The state is not normalized and its norm
$P(\beta)=\mbox{Tr}_{B}\left[\tilde{\rho}_{\rm B}(\beta)\right]$
gives the probability density of finding the outcome $\beta$. The
probability that the measurement outcome $\beta$ falls into the
set $\Sigma$ then reads
$P_{\Sigma}=(1/\pi)\int_{\Sigma}P(\beta)d^{2}\beta$ and the
corresponding conditionally prepared normalized density matrix is
then given by $\rho_{\Sigma}=\frac{1}{\pi
P_{\Sigma}}\int_{\Sigma}\tilde{\rho}_{B}(\beta)d^{2}\beta$. Making
use the formula for the Wigner function of the state
$\rho_{\Sigma}$ of the form
$W_{\Sigma}(0)=(1/\pi)\mbox{Tr}_{B}\left[\rho_{\Sigma}(-1)^{n}\right]$
\cite{Royer_77} we find the output Wigner function in the origin
can be expressed as the integral
\begin{equation}\label{WSigma0}
W_{\Sigma}(0)=\frac{1}{\pi
P_{\Sigma}}\int_{\Sigma}W_{\tilde{\rho}_{B}(\beta)}(0)d^{2}\beta,
\end{equation}
of the Wigner function in the origin
$W_{\tilde{\rho}_{B}(\beta)}(0)$ of the state (\ref{rhobeta}).

First, we will analyze the most simple case when $\rho_{\rm
in}=|1\rangle_{\rm in}\langle 1|$. Then the probability density
$P(\beta)$ reads
\begin{equation}\label{Pbeta}
P(\beta)=_A\!\langle
1|D_{A}(\beta^{\ast})\rho_{A}D_{A}^{\dag}(\beta^{\ast})|1\rangle_{A},
\end{equation}
where $\rho_{A}=\mbox{Tr}_{B}\rho_{AB}$ is the reduced state of
mode $A$. The reduced state is a thermal state with mean number of
thermal photons equal to $\langle n\rangle=(V-1)/2$, where
$V=V_{\rm an}+V_{\rm sq}$ and the probability density is therefore
an overlap of the displaced thermal state with the Fock state
$|1\rangle$. Expressing the overlap in terms of the Wigner
functions and performing the needed integration we arrive at the
probability density of the form
\begin{equation}\label{Pbetan}
P(\beta)=\frac{\langle n\rangle}{\left(1+\langle
n\rangle\right)^2}\left[1+\frac{|\beta|^2}{\langle
n\rangle\left(1+\langle
n\rangle\right)}\right]e^{-\frac{|\beta|^2}{1+\langle n\rangle}}.
\end{equation}
For the sake of computational simplicity we will assume the set
$\Sigma$ to be a square in the plane
$\left[\bar{x}_{u},\bar{p}_{v}\right]$ of measurement outcomes
centered in the origin with sides of length $2a$ parallel with the
coordinate axes. Integration of the probability density
(\ref{Pbetan}) over the square then yields the success probability
$P_{\Sigma}$ the explicit form of which in terms of the error
function is given by Eq.~(\ref{PSigmaexplicit}) in the Appendix.
For the measurement outcome $\beta$ the output state
(\ref{rhobeta}) attains the form
\begin{equation}\label{rhobeta1}
\tilde{\rho}_{\rm B}(\beta)=D_{B}(G\beta)_{A}\!\langle
1|D_{A}(\beta^{\ast})\rho_{AB}
D_{A}^{\dag}(\beta^{\ast})|1\rangle_{A}D_{B}^{\dag}(G\beta),
\end{equation}
where we used the relation $_{\rm in}\langle 1|\Pi_{{\rm in}
A}(\beta)|1\rangle_{\rm
in}=D^{\dag}_{A}(\beta^{\ast})|1\rangle_{A}\langle
1|D_{A}(\beta^{\ast})$. The Wigner function in the origin of the
state (\ref{rhobeta1}) needed to calculate the Wigner function in
the origin (\ref{WSigma0}) then can be calculated from the overlap
formula
\begin{equation}\label{Wrhobeta}
W_{\tilde{\rho}_{B}(\beta)}(0)=2\pi\int_{-\infty}^{+\infty}
W_{AB}(\xi_{A}-\bar{\xi}_{A},-\bar{\xi}_{B})W_{A}(\xi_{A})d\xi_{A},
\end{equation}
where $\xi_{A}=(x_{A},p_{A})^{T}$,
$\bar{\xi}_{A}=(\bar{x}_{u},-\bar{p}_{v})^{T}$,
$\bar{\xi}_{B}=G(\bar{x}_{u},\bar{p}_{v})^{T}$, $W_{A}(\xi_{A})$
is the Wigner function of the Fock state $|1\rangle$ given in
Eq.~(\ref{Win}) and
\begin{equation}\label{WAB}
W_{AB}(\xi)=\frac{1}{4\pi^{2}V_{\rm sq}V_{\rm
an}}e^{-\xi^{T}\gamma_{AB}^{-1}\xi},
\end{equation}
where $\xi=(x_{A},p_{A},x_{B},p_{B})^{T}$ and $\gamma_{AB}$ is the
covariance matrix (CM) of the shared state $\rho_{AB}$ of the form:
\begin{eqnarray}\label{gammaAB}
\gamma_{AB}=\left(\begin{array}{cc}
V\openone & C\sigma_{z} \\
C\sigma_{z} & V\openone\\
\end{array}\right),
\end{eqnarray}
where $C=V_{\rm an}-V_{\rm sq}$. Notice that we use the notation in which the CM of a vacuum state is equal
to $\gamma_{\rm vac}=\openone$ and that the matrix (\ref{gammaAB}) is a legitimate CM of
a quantum state. Namely, denoting its submatrices as $A=B=V\openone$
and $D=C\sigma_{z}$ one can show easily using the Heisenberg uncertainty relations $V_{\rm sq}V_{\rm an}\geq1/4$
that the matrix (\ref{gammaAB}) satisfies the necessary and sufficient conditions for a matrix to be a CM of
a quantum state given by the inequalities $A,B>0$, $\mbox{det}A+\mbox{det}B+2\mbox{det}D\leq 1+\mbox{det}\gamma_{AB}$ and
$2\sqrt{\mbox{det}A\mbox{det}B}+\left(\mbox{det}D\right)^2\leq\mbox{det}\gamma_{AB}+\mbox{det}A\mbox{det}B$ \cite{Pirandola_09}.

Performing now the integration in Eq.~(\ref{Wrhobeta}) using Eqs.~(\ref{WAB}) and (\ref{gammaAB})
and substituting the obtained formula into Eq.~(\ref{WSigma0}) we
finally get after integration over $\beta$ the sought Wigner
function in the origin $W_{\Sigma}(0)$ that is explicitly given by
formula (\ref{WSigma0explicit}) in the Appendix.

\subsection{Conditioning on the outcome $\beta=0$}

\begin{figure}[h]
\centerline{\psfig{width=8cm,angle=0,file=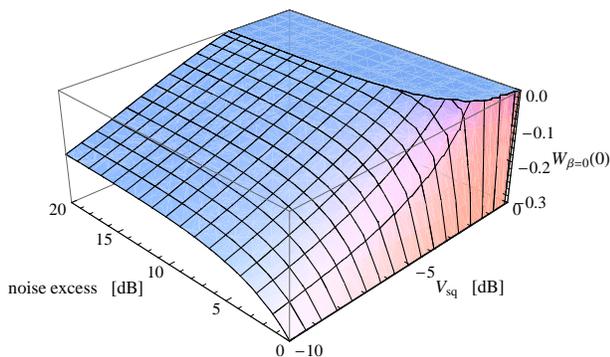}}
\caption{Output Wigner function in the origin versus the squeezed
variance $V_{\rm sq}=\langle\left(\Delta
p_{A}\right)^2\rangle=\langle\left(\Delta x_{B}\right)^2\rangle$
and the noise excess in the anti-squeezed quadratures $V_{\rm
an}=\langle\left(\Delta x_{A}\right)^2\rangle=\langle\left(\Delta
p_{B}\right)^2\rangle$ for conditional teleportation of the Fock
state $|1\rangle$ where we accept only the result $\beta=
(\bar{x}_{u}+i\bar{p}_{v})/\sqrt{2}=0$. }\label{figure9}
\end{figure}

Let us first analyze the most simple case when we accept only the
measurement outcome $\beta=0$. Then using Eq.~(\ref{rhobeta1}) we
obtain the normalized output state in the form $_{A}\!\langle
1|\rho_{AB}|1\rangle_{A}/P(0)$, where $P(0)=\langle
n\rangle/\left(1+\langle n\rangle\right)^2$ is obtained from
Eq.~(\ref{Pbetan}). The Wigner function in the origin of the state
can be derived with the help of the formula (\ref{Wrhobeta}) in
the form
\begin{equation}\label{Wbeta0}
W_{\beta=0}(0)=-\frac{(V-4V_{\rm sq}V_{\rm an})}{\pi
\left(V-1\right)}\left(\frac{V+1}{V+4V_{\rm sq}V_{\rm
an}}\right)^{2}
\end{equation}
and it is depicted in Fig.~\ref{figure9} as a function of the
squeezed variance $V_{\rm sq}$ and the noise excess defined as the
sum of the anti-squeezed quadrature in decibels and squeezed
quadrature in decibels.
The figure shows again that for zero noise excess the Fock state
$|1\rangle$ is perfectly teleported for arbitrarily small nonzero
squeezed variance $V_{\rm sq}$ as it is also apparent from
Eq.~(\ref{psibeta}). For a nonzero noise excess there is a
threshold squeezed variance $V_{\rm th}$ one has to overcome in
order to have a negative output Wigner function in the origin. The
threshold squeezed variance can be determined from the condition
$V-4V_{\rm sq}V_{\rm an}=0$. We calculated threshold squeezed
variance for several values of the noise excess in
Table~\ref{table1}. The calculated values indicate that the
threshold squeezed variance increases very slowly with increasing
noise excess and approaches the limit value $V_{{\rm
th},\infty}=1/4$ corresponding to -3 dB in the limit of infinitely
large noise excess. Thus in the limit case of post-selection of
the measurement outcome $\beta=0$ the teleportation of a negative
Wigner function is strongly tolerable to the noise excess.
\begin{table}[ht]
\caption{Threshold squeezed variance $V_{\rm th}$ for a given
noise excess if we post-select the measurement outcome $\beta=0$.}
\centering
\begin{tabular}{| c | c | c | c | c | c |}
\hline noise excess [dB] & 1 & 2 & 3 & 4 & 5 \\
\hline $V_{\rm th}$ [dB] & -1.62 & -2.06 & -2.32 & -2.49 & -2.62 \\
\hline
\end{tabular}\label{table1}
\end{table}

\subsection{Finite post-selection interval}

Next we will focus on the case of a finite post-selection
interval. The success probability and the Wigner function in the
origin are given by Eqs.~(\ref{PSigmaexplicit}) and
(\ref{WSigma0explicit}) in the Appendix and they are displayed in
Figs.~\ref{figure10} and \ref{figure11} for $a=0.3$ and unity gain
regime ($G=1$). Inspection of the graph in Fig.~\ref{figure11}
reveals that it is just a displaced graph in Fig.~\ref{figure9}
along the $z$ axis. In other words, conditioning on measurements
outcomes from a finite post-selection interval leads to the
uniform reduction of the value of the output Wigner function in
the origin in comparison with the case when we accept only the
outcome $\beta=0$. This naturally entails emergence of a nonzero
threshold on squeezing that has to be overcome in order to
successfully teleport a negative Wigner function. The threshold
still increases slowly with increasing noise excess as is apparent
from the Table~ \ref{table2}.

\begin{table}[ht]
\caption{Threshold squeezed variance $V_{\rm th}$ for a given
noise excess for a finite post-selection interval.} \centering
\begin{tabular}{| c | c | c | c | c | c | c |}
\hline noise excess [dB] & 0 & 1 & 2 & 3 & 4 & 5 \\
\hline $V_{\rm th}$ [dB]
& -1.13 & -1.77 & -2.12 & -2.35 & -2.51 & -2.63 \\
\hline
\end{tabular}\label{table2}
\end{table}

\begin{figure}[h]
\centerline{\psfig{width=7cm,angle=0,file=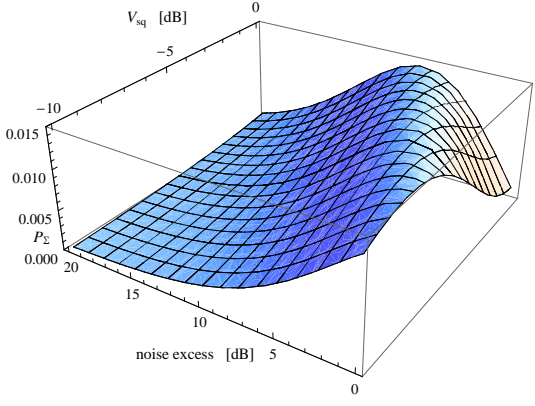}}
\caption{Success probability $P_{\Sigma}$ versus the squeezed
variance $V_{\rm sq}=\langle\left(\Delta
p_{A}\right)^2\rangle=\langle\left(\Delta x_{B}\right)^2\rangle$
and the noise excess in the anti-squeezed quadratures $V_{\rm
an}=\langle\left(\Delta x_{A}\right)^2\rangle=\langle\left(\Delta
p_{B}\right)^2\rangle$ for conditional unity gain teleportation of
the Fock state $|1\rangle$ where we accept only the outcomes of
the Bell measurement $[\bar{x}_{u},\bar{p}_{v}]$ that fall into
the square centered in the origin with the side $2a$ where
$a=0.3$. }\label{figure10}
\end{figure}
\begin{figure}[h]
\centerline{\psfig{width=8.5cm,angle=0,file=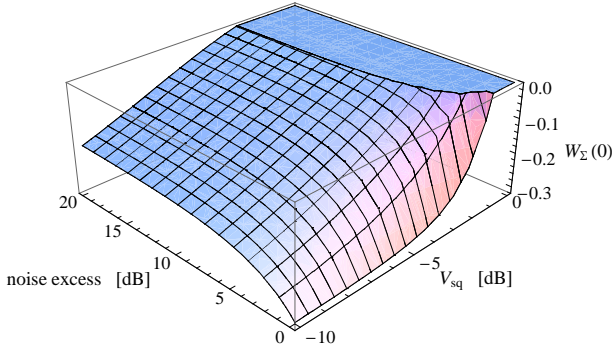}}
\caption{Output Wigner function in the origin versus the squeezed
variance $V_{\rm sq}=\langle\left(\Delta
p_{A}\right)^2\rangle=\langle\left(\Delta x_{B}\right)^2\rangle$
and the noise excess in the anti-squeezed quadratures $V_{\rm
an}=\langle\left(\Delta x_{A}\right)^2\rangle=\langle\left(\Delta
p_{B}\right)^2\rangle$ for conditional unity gain teleportation of
the Fock state $|1\rangle$ where we accept only the outcomes of
the Bell measurement $[\bar{x}_{u},\bar{p}_{v}]$ that fall into
the square centered in the origin with side $2a$ where
$a=0.3$.}\label{figure11}
\end{figure}

\subsection{Attenuated single-photon state}

Finally, we will discuss the situation when we have the attenuated
single-photon state at the input, i.e., $\rho_{\rm
in}=\rho_{\eta}$, where $\rho_{\eta}$ is defined in
Eq.~(\ref{rhoeta}) and we condition on the measurement outcome
$\beta=0$.

\begin{figure}[h]
\centerline{\psfig{width=6.5cm,angle=0,file=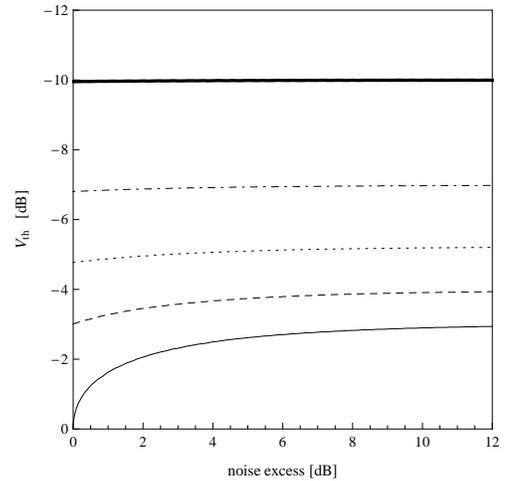}}
\caption{Threshold squeezed variance $V_{\rm th}$ versus the noise
excess for conditional teleportation of the input state
$\eta|1\rangle\langle 1|+(1-\eta)|0\rangle\langle 0|$ for $\eta=1$
(solid curve), $\eta=0.9$ (dashed curve), $\eta=0.8$ (dotted
curve), $\eta=0.7$ (dash-dotted curve) and $\eta=0.6$ (solid thick
curve), where we accept only the result $\beta=
(\bar{x}_{u}+i\bar{p}_{v})/\sqrt{2}=0$.}\label{figure12}
\end{figure}

In this case the probability density reads
\begin{equation}\label{Petabeta0}
P^{(\eta)}(\beta)=\eta P(\beta)+(1-\eta)P^{(0)}(\beta),
\end{equation}
where $P(\beta)$ is given in Eq.~(\ref{Pbetan}) and
\begin{eqnarray}\label{Peta0}
P^{(0)}(\beta)&=&_A\!\langle
0|D_{A}(\beta^{\ast})\rho_{A}D_{A}^{\dag}(\beta^{\ast})|0\rangle_{A}\nonumber\\
&=&\frac{e^{-\frac{|\beta|^{2}}{1+\langle n\rangle}}}{1+\langle
n\rangle}.
\end{eqnarray}
For $\beta=0$ we get, in particular,
$P^{(\eta)}(0)=2\left(V+1-2\eta\right)/\left(V+1\right)^2$.
Further, conditioned on the measurement outcome $\beta=0$ we get
the unnormalized output state in the form:
\begin{eqnarray}\label{rhoetabeta0}
\tilde{\rho}^{(\eta)}_{\rm B}(0)=\eta_{A}\!\langle 1|\rho_{AB}
|1\rangle_{A}+(1-\eta)_{A}\!\langle 0|\rho_{AB} |0\rangle_{A}.
\end{eqnarray}
Calculating finally the Wigner function in the origin of the
normalized state $\tilde{\rho}^{(\eta)}_{\rm B}(0)/P^{(\eta)}(0)$
we obtain
\begin{equation}\label{Wetabeta0}
W^{(\eta)}_{\beta=0}(0)=\frac{4V_{\rm sq}V_{\rm
an}+V\left(1-2\eta\right)}{\pi
\left(V+1-2\eta\right)}\left(\frac{V+1}{V+4V_{\rm sq}V_{\rm
an}}\right)^{2}.
\end{equation}
Hence, one can calculate the threshold squeezed variance from the
condition $4V_{\rm sq}V_{\rm an}+V\left(1-2\eta\right)=0$.
Expressing the anti-squeezed quadrature as $V_{\rm
an}=(1/2)(N/V_{\rm sq})$, where $N$ stands for the noise excess
($N=1/2$ for zero noise excess), we get the threshold squeezed
variance in the form:
\begin{equation}\label{Vthetabeta0}
V_{\rm
th}=\frac{2N-\sqrt{4N^2-2N(2\eta-1)^2}}{2\left(2\eta-1\right)}.
\end{equation}
The threshold variance in dependence of the noise excess is
plotted in Fig.~\ref{figure12} for several values of the
probability $\eta$. The threshold squeezing dramatically increases
with decreasing probability of the Fock state $|1\rangle$ but for
realistic values of the probability $\eta$ in the interval
$0.6<\eta<0.7$ it still remains in the region of achievable
squeezing levels. As in the previous cases saturation effect is
observed and the threshold squeezed variance approaches $V_{{\rm
th},\infty}=(2\eta-1)/4$ in the limit of infinitely large noise
excess. In order to illustrate the tolerance of the teleportation
of a negative Wigner function to noise excess even for mixed input
states we will consider again the example of the input state
$\rho_{\eta}$ with $\eta=0.6304$ discussed above and 4 dB noise
excess. Using Eq.~(\ref{Vthetabeta0}) we then get the threshold
squeezed variance $V_{\rm th}$ equal to -8.82 dB and it approaches
$V_{{\rm th},\infty}$ equal to -8.85 dB with increasing noise
excess.

\section{Conclusions and Discussion}\label{sec_4}

We studied teleportation of a pure single-photon Fock state and a
mixed attenuated single-photon Fock state by the standard
continuous-variable teleportation protocol \cite{Braunstein_98}.
We optimized analytically the gain of the teleportation such in
order to minimize the output Wigner function in the origin. For
the single-photon Fock state we found that an arbitrarily weak
squeezing used to create the shared entangled state is sufficient
for successful teleportation of the negative value of its Wigner
function in the origin. For an attenuated single-photon state we
have shown that there is a strict bound on the squeezing that has
to be overcome in order to have the output Wigner function in the
origin negative. In both cases the negative value of the output
Wigner function in the origin can be increased by using a
conditional teleportation with a reasonably high success rate.
However, in the case of the attenuated single-photon state, the
bound on squeezing one has to surpass to observe negative output
Wigner function is the same for both unconditional and
conditional teleportation and its initial value can be reached
only asymptotically in the limit of a high squeezing. Finally, we
took into account noise excess in the anti-squeezing and we have shown
that conditional teleportation of a negative Wigner function exhibits
strong tolerance to the noise excess.

Now, let us discuss the consequences of our observations for the
quantum computation and the quantum communication. For quantum
computation, the quantum teleportation was considered as a
possible scenario how to make {\em deterministic} operation on an
unknown arbitrary quantum state. A basic requirement is to be able
to preserve negativity of the Wigner function through the
operation, in our simplest case through the teleportation.
Therefore, we can take the preservation of the negativity as a
necessary condition for the quantum gate performance. We observed
that to preserve the negativity of the Winger function, either the
input state has to be very close to a single-photon state, or for
the imperfect single-photon state (the attenuated version), an
extremely high squeezing is required. If the negativity of the
Wigner function should be preserved (for example, at $95\%$ of its
original value), the requirement is even more demanding. Clearly,
it is very important to protect the input state against even the
loss, since it substantially increases the demand on the required
squeezing for the teleportation. In summary, the cost (squeezing
required to prepare entangled state in the teleportation) is quite
high to implement the operation in this measurement-induced way.
It can stimulate further increasing of the squeezing in the
experiments, but this resource seems to be also practically
limited.

On the other hand, for quantum communication with the repeaters,
the preservation of negativity of the Wigner function through the
Gaussian teleportation seems to be just a reasonable condition to
efficiently extend the quantum key distribution between the two
distant repeaters. Further, it is enough to implement the {\em
conditional} teleportation, since the key distribution is anyway
probabilistic protocol. The required minimal squeezing to keep the
negativity of the Wigner function is practically not changed by
the post-selections. However, for almost perfect single-photon
state, the threshold squeezing is low and the input negativity of
the Wigner function can be archived for the experimentally
feasible values of squeezing (up to -10 dB of squeezing).
Advantageously, for the strongly attenuated single-photon state,
the post-selection improves the value of the negativity up to a
maximum for the given squeezing used to produce entanglement.
However, to reach the original value, the post-selection has to be
combined with the enhancement of the squeezing. In summary, the
cost (squeezing required to prepare entangled state in the
teleportation) is lower for the communication application of the
teleportation of highly non-classical states, if the negativity of
the Wigner function is already preserved by teleportation.

Finally, we want to point out one important issue. Since the
teleportation with finite squeezing always lowers the negativity
of the Wigner function, any next teleportation will transfer
already an imperfect version of the highly non-classical state and
the squeezing threshold (or squeezing required to almost maintain
the negativity) will become more demanding for the implementation.
It means that small imperfections or errors are actually amplified
through multiple Gaussian teleportations and any correction
mechanism (quantum error correction or quantum repeater) has to be
applied very frequently and very efficiently to keep the threshold
squeezing in the feasible range to be able to transmit the
negativity of the Wigner function.
\acknowledgments

The research has been supported by the research projects
``Measurement and Information in Optics,'' (MSM 6198959213),
Center of Modern Optics (LC06007) of the Czech Ministry of
Education, project Czech-Japan project ME10156 (MIQIP) of MSMT and
the project of GACR No. 202/08/0224. We also acknowledge the
financial support of the EU under FET-Open project COMPAS
(212008).
\appendix*
\section{}

The success probability $P_{\Sigma}$ is given by:
\begin{eqnarray}\label{PSigmaexplicit}
P_{\Sigma}=\operatorname{erf}\left(b\right)\left[\operatorname{erf}\left(b\right)-\frac{2be^{-b^2}}{\sqrt{\pi}\left(1+\langle
n\rangle\right)}\right],
\end{eqnarray}
where $b=a/\sqrt{2\left(1+\langle n\rangle\right)}$ and
$\operatorname{erf}\left(z\right)=(2/\sqrt{\pi})\int_{0}^{z}e^{-t^2}dt$
is the error function.

The conditional output Wigner function in the origin
$W_{\Sigma}(0)$ reads
\begin{widetext}
\begin{eqnarray}\label{WSigma0explicit}
W_{\Sigma}(0)=\frac{\operatorname{erf}\left(a\sqrt{q}\right)}{4\pi
P_{\Sigma}V_{\rm sq}V_{\rm an}\alpha
q}\left[\left(\frac{2}{\alpha}+\frac{2\delta^2}{\alpha^2q}-1\right)\operatorname{erf}\left(a\sqrt{q}\right)-\frac{4\delta^2a}{\alpha^2\sqrt{\pi
q}}e^{-qa^2}\right],
\end{eqnarray}
\end{widetext}
where
\begin{widetext}
\begin{eqnarray*}
q&=&\frac{V\left(1+G^{2}\right)-2CG+G^2}{V+4V_{\rm sq}V_{\rm
an}},\quad \alpha=\frac{V+4V_{\rm sq}V_{\rm an}}{4V_{\rm sq}V_{\rm
an}},\quad \delta=\frac{CG-V}{4V_{\rm sq}V_{\rm an}}.
\end{eqnarray*}
\end{widetext}

\end{document}